\documentclass[aps,prl,twocolumn,superscriptaddress]{revtex4-1}
\usepackage{amsmath,bbm,amssymb}
\usepackage{xcolor,graphics,graphicx}
\usepackage[caption=false]{subfig}
\usepackage[export]{adjustbox}

\newcommand{\be}{\begin{equation}}
\newcommand{\ee}{\end{equation}}
\newcommand{\bea}{\begin{align}}
\newcommand{\eea}{\end{eqnarray}}
\newcommand{\id}{\mathbbm{1}}

\newcommand{\sig}{{\boldsymbol{\sigma}}}

\begin{document}


\title{Gaussian Thermal Operations and the Limits of Algorithmic Cooling}


\author{A. Serafini}
\affiliation{Department of Physics \& Astronomy, University College London, Gower Street, WC1E 6BT, London, United Kingdom}
\author{M. Lostaglio}
\affiliation{ICFO - Institut de Ci{\`e}ncies Fot{\`o}niques, The Barcelona Institute of Science and Technology, 08860 Castelldefels, Spain}
\author{S. Longden}
\affiliation{Department of Physics \& Astronomy, University College London, Gower Street, WC1E 6BT, London, United Kingdom}
\author{U. Shackerley-Bennett}
\affiliation{Department of Physics \& Astronomy, University College London, Gower Street, WC1E 6BT, London, United Kingdom}
\author{C.-Y. Hsieh}
\affiliation{ICFO - Institut de Ci{\`e}ncies Fot{\`o}niques, The Barcelona Institute of Science and Technology, 08860 Castelldefels, Spain}
\author{G. Adesso}
\affiliation{School of Mathematical Sciences and Centre for the Mathematics and Theoretical Physics of Quantum Non-Equilibrium Systems,
University of Nottingham, University Park Campus, Nottingham NG7 2RD, United Kingdom}


\date{\today}

\begin{abstract}

The study of thermal operations 
allows one to investigate the ultimate possibilities of quantum states and of nanoscale thermal machines. Whilst fairly general, these results 
typically do not apply to continuous variable systems and do not take into account that, in many practically relevant settings, system-environment interactions are effectively bilinear. 
Here we tackle these issues by focusing on Gaussian quantum states and channels.
 We provide a complete characterisation of the most general Gaussian thermal operation acting on an arbitrary number of bosonic modes,
 which turn out to be all embeddable in a Markovian dynamics, 
 and derive necessary and sufficient conditions for state transformations under such operations in the single-mode case, encompassing states with nonzero coherence in the energy eigenbasis (i.e., squeezed states). Our analysis leads to a no-go result for the technologically relevant task of algorithmic cooling: We show that it is impossible to reduce the entropy of a system coupled to a Gaussian environment below its own or the environmental temperature, by means of a sequence of Gaussian thermal operations interspersed by arbitrary (even non-Gaussian) unitaries. These findings establish fundamental constraints on the usefulness of Gaussian resources for quantum thermodynamic processes.

\end{abstract}


\maketitle


\noindent {\em Introduction and Summary --} The past few years have witnessed 
a resurgence of studies into the thermodynamics of quantum systems \cite{thermobook}, 
which have lent novel
insight into the nature of thermodynamic relations, as
well as into the role of thermodynamic quantities such as temperature, entropy and work \cite{goold2016role, vinjanampathy2016quantum}, set against the practical backdrop of realising  
superior thermal machines operating in the quantum regime~\cite{alicki2018introduction}. 

A key ingredient to any attempt to analyse these questions beyond the limited scope of a specific model is the characterisation of a class of ``thermal operations'', i.e., of operations that can be realised with the aid of the surrounding environment \cite{janzing2000thermodynamic, brandao2011resource} (for reviews, see \cite{lostaglio2018thermodynamic, ng2018resource}). 
Whilst the frameworks resulting from this approach may yield significant wisdom concerning the ultimate limitations
that constrain thermal scenarios, they are at times fraught by a certain `lack of realism', in that they include interaction Hamiltonians which are not necessarily encountered in practice. Also, they are limited to finite-dimensional settings.
It is therefore desirable to single out and characterise subclasses of thermal operations with direct practical relevance.

To this aim, this paper shall consider the subclass of \emph{Gaussian thermal operations} (GTOs), i.e., the class of operations on continuous variable systems obtained by considering energy-preserving bilinear interaction Hamiltonians between the system and a thermal environment.
This subclass is extremely relevant in practice, given that quadratic Hamiltonians, which generate Gaussian unitaries, 
are very common and that system-bath interactions are often linear or may be linearised, especially in quantum optics and analogous set-ups. 
Indeed, various experimental platform relevant to quantum thermodynamics operate in the Gaussian regime. 
Examples include cavity optomechanics \cite{brunelli15,mari15}, 
Bose-Einstein condensates loaded into cavities~\cite{brunelli2018experimental}, 
ions in harmonic traps \cite{huber08,shuoming15}. In view of the same practical reasons, work extraction, storage and fluctuations~\cite{brown16, friis2018precisionwork, deffner2008nonequilibrium, singh2019quantum}, entropy production  \cite{belenchia2019entropy}, heat transport \cite{dhar2008heat}, thermometry~\cite{correa2016low} and fluctuation-dissipation theorems~\cite{mehboudi2019linear} have been investigated in Gaussian scenarios. 
The capabilities of Gaussian operations in other, not necessarily thermodynamical, settings 
are also being considered \cite{lami18, yadin2018operational}. 



The other defining feature of GTOs, alongside Gaussianity, 
is energy-preservation. As we will prove, this feature implies that GTOs may be physically reproduced
through operations corresponding, in the optical picture, to passive optical elements (i.e., semi-reflectant mirrors and dielectric plates), even in presence of thermal noise (which can be represented as a beam splitter coupling the input with a thermal mode). Alternately, dynamics equivalent to GTOs can be obtained by contact with a Markovian thermal reservoir (giving rise to the so-called quantum optical master equation, see \cite{wallsmil,bucca}), or in coupled resonant cavities or optomechanical systems with negligible counter-rotating terms \cite{sala2018exploring}. 

In this paper, we shall achieve a compact, constructive characterisation of the most general GTO on any number of modes. We shall see that such a characterisation becomes particularly simple for systems with non-degenerate eigenfrequencies, where it can be cast as a single-mode property. 
We shall then derive necessary and sufficient conditions for state transformation on single-mode systems and then proceed
to analyse the possibilities offered by algorithmic cooling in the Gaussian regime,
through alternating GTOs and unitaries. We will prove that, at variance with the finite-dimensional case \cite{alhambra2018heat}, in the absence of ancillas no such strategy can cool the system below the environmental temperature. Sideband-like strategies involving high-frequency ancillas or higher order interactions are necessary to such an aim. 
Also in view of the ubiquity of Gaussian evolutions as a complete toolbox for quantum technologies \cite{adesso14} and in the modelling of open quantum systems of harmonic systems,
the fundamental limitations to cooling techniques we will establish in the Gaussian regime possess a direct practical interest. 

%
\smallskip


\noindent {\em Gaussian systems --}  We will consider bosonic continuous variables encoded into vectors of
self-adjoint operators $\hat{\bf r} = (\hat{x}_1,\hat{p}_1,\ldots,\hat{x}_n,\hat{p}_n)^{\sf T}$
obeying the canonical commutation relations
$[\hat{\bf r},\hat{\bf r}^{\sf T}]= i\Omega$, where the commutators are taken between all pairs of elements of $\hat{\bf r}$
(as in an outer product)
and form the non-degenerate, antisymmetric symplectic form $\Omega$, with
$\Omega  = \Omega_1^{\oplus n}$ and
$\Omega_1 = \left( \begin{smallmatrix} 0 & 1 \\  -1 & 0 \end{smallmatrix} \right)$ \cite{bucca}.
A second-order Hamiltonian $\hat{H}$ is
one that may be written as a second-order polynomial of $\hat{\bf r}$:
$\hat{H}=\frac12 (\hat{\bf r}-{\bf d})^{\sf T} H (\hat{\bf r} - {\bf d})$
for a symmetric Hamiltonian matrix $H$ and a real vector ${\bf d}$. Gaussian states are then defined as the ground and
(Gibbs) thermal states of second-order Hamiltonians,
and are completely characterised by a vector of first
moments ${\bf r} = \langle \hat{\bf r} \rangle $ and the covariance matrix (CM)
$\sig = \langle \{(\hat{\bf r}-{\bf r}),(\hat{\bf r}-{\bf r})^{\sf T}\} \rangle$ where, again, the anticommutators $\{\cdot,\cdot\}$
are taken between each pair of operator entries to form the symmetric, real matrix $\sig$, satisfying 
$\sig+i\Omega\ge 0$ \cite{bucca}.
Gaussian unitary operations -- ones that map Gaussian states into Gaussian states --
are those generated by second-order Hamiltonians and
admit a symplectic representation: their action on the second moments may be written as
$\sig \mapsto S \sig S^{\sf T}$, where $S\in Sp_{2n,{\mathbbm R}}$ (i.e., $S$ is such that $S \Omega S^{\sf T}=\Omega$).
It is well known that any positive-definite real matrix $P$ may be put into `normal modes' by congruence with
a symplectic transformation: $\exists \, S \; : \; SPS^{\sf T} = \oplus_{j=1}^{n} \nu_j\id_2  $, where the
$\nu_j$'s are the `symplectic eigenvalues' of $P$; if $P$ is a Hamiltonian
matrix, the quantities $\nu_j$ represent the eigenfrequencies of $P$ (the frequencies of its normal modes).
In the case of the CM of a quantum state, one has $\nu_j\ge 1$ (an expression of the uncertainty principle).
Bear in mind that the spectrum of a Gaussian state is entirely determined by its symplectic eigenvalues
and that tensor products at the Hilbert space level translate into direct sums in the Gaussian and phase space descriptions.

In the following, a major role will be played by the set of orthogonal symplectic transformations,
for which $S S^{\sf T}=\id$ (also known as ``compact'', or ``passive''
transformations, as they do not require any source of energy in standard
optical implementations).
Further specific notation will prove convenient:
we shall adopt the shorthand notation $S[\sig] = S \sig S^{\sf T}$
and the symbol ${\rm Tr}_b$ to denote partial tracing of the bath's degrees of freedom in the phase space,
which just corresponds to pinching out the relevant part of a CM, discarding the rest.




Let us also recall that the most general deterministic Gaussian CP-map, 
obtained by letting the system interact with an environment in a Gaussian state through a quadratic interaction Hamiltonian,
is characterised, up to arbitrary displacements of the first moments, by the mapping $\sig \mapsto
X\sig X^{\sf T} +Y$, with $Y \ge - i X\Omega X^{\sf T} + i\Omega$ \cite{bucca}.
The first aim of this paper will be characterising the subclass of deterministic Gaussian CP-maps that are also thermal.
A particularly relevant class of single-mode channels, which will play a prominent role in what follows, 
is the so-called `phase-covariant' ones, where
$X= x\id_2$ and $Y=y \id_2$, with $y\ge |1-x^2|$ \cite{bucca,giovannetti2014ultimate},
(throughout the paper, the symbol $\id_d$ denotes the identity matrix in dimension $d$).\smallskip

\noindent {\em The class of Gaussian thermal operations --} Given a second-order system
Hamiltonian $\hat{H}_s$ and an inverse temperature $\beta = 1/(kT)$
(where $k$ is the Boltzmann's constant and $T$ is the environment's temperature), 
we shall define GTOs as the operations obtained by:
\begin{itemize}
\item Preparing an environmental ancilla with arbitrary second-order Hamiltonian $\hat{H}_b$ in the Gibbs state ${\rm e}^{-\beta \hat{H}_b}/{\rm Tr}\left[{\rm e}^{-\beta \hat{H}_b}\right]$.
\item Letting system and bath interact through an energy preserving Gaussian unitary $\hat{U}_{I}$ such that \mbox{$[\hat{U}_I, \hat{H}_s + \hat{H}_b]=0$}.
\end{itemize}
The maps above arise naturally through contact with thermal reservoirs where the interactions are well described by polynomials of order two in the canonical operators, whose importance has been already remarked.
Note that all energy preserving Gaussian unitaries can be written as $\hat{U}_I =  e^{i \hat{H}_I t}$ for some $t \geq 0$ and $\hat{H}_I$ a Hamiltonian of order two in the canonical operators satisfying $[\hat{H}_I, \hat{H}_s + \hat{H}_b] = 0$.
Note also that the definition above coincides with the well-established definition of thermal operations 
\cite{janzing2000thermodynamic, brandao2011resource}, once the restrictions to second-order operations are lifted.

Arbitrary $\hat{H}_s$, $\hat{H}_b$ and $\hat{H}_I$ of order two are parametrised by the symmetric Hamiltonian matrices
$H_s$, $H_b$ and $H_{I}$ and the vectors ${\bf d}_s$, ${\bf d}_b$ and ${\bf d}_I$. We will further restrict the Hamiltonian matrices of system and environment to be strictly positive. Hamiltonian matrices with negative eigenvalues correspond to Hamiltonian operators that are not bounded from below, and thus do not even admit a well-defined Gibbs state, so their exclusion is not a restriction.
Positive semi-definite, but not strictly positive, Hamiltonian matrices correspond to a set of measure zero within the Gaussian realm, with Gibbs states that are not regular, trace-class Gaussian states and thus do not give rise to Gaussian CP-maps. It might still be possible to obtain legitimate operations from non-positive system Hamiltonians, but we shall disregard such peculiar cases in this treatment.

First-order terms in the interaction Hamiltonian generate displacements (shifts in the first-moment vector ${\bf r}$). Since no first-order term commutes with a strictly positive quadratic Hamiltonian (linear displacements do affect the energy of trapped systems), displacements must be severely limited if they are to  give rise to thermal operations. 
Rather than complicating our treatment with the inclusion of first-order terms,
which do not add anything conceptually remarkable,
we defer such a discussion to the Supplemental Material (SM) \cite{SM}, 
and set all first order terms ${\bf d}_{s}$, ${\bf d}_b$ and ${\bf d}_{I}$ to zero to present our main results.\smallskip

\noindent \emph{Simulating Gaussian thermalisations --} Within the above restrictions, a GTO generally involves an arbitrary number of bath modes, as well as an arbitrary sequence of second order interactions between these and the system modes. A crucial question is then if there exists a simpler protocol able to reproduce every Gaussian thermalisation with less extensive resources. Our first main result answers this question in the affirmative, presenting a very simple scheme able to simulate exactly a general GTO (recall the shorthand notation whereby symplectics act by congruence):

\smallskip

\noindent {\bf Theorem 1 -- Characterisation of GTOs.} {\em Let $\hat{H}_s = \frac12
\hat{\bf r}^{\sf T}H_s \hat{\bf r}$ be a system Hamiltonian with normal form $
\bigoplus_l \omega_l \id_{2n_l} = S^{-1}H_s S^{\sf T-1}$, where $n_l\in {\mathbbm N}$ is the mode degeneracy of
the eigenfrequency
$\omega_l$ and $S\in Sp_{2n,{\mathbbm R}}$ for $n=\sum_l n_l$. 
The class of GTOs at background inverse temperature $\beta$ act on the system CM $\sig$ as
\be
\sig \mapsto S\left[ \oplus_{l} W_l \circ \Phi_l \circ Z_l \left[S^{-1}[\sig]\right] \right]
\; ,
\ee
where the direct sum runs over the distinct eigenfrequencies and, setting $\nu_l = \frac{{\rm e}^{\beta\omega_l}+1}{{\rm e}^{\beta\omega_l}-1}$:
\begin{enumerate}
\item Each $\Phi_l$ are phase-covariant CP maps \cite{giovannetti2014ultimate}, 
acting on the $l-$th eigenfrequency space as $\Phi_l(\sig) =  X_l \sig X_l^{\sf T} + Y_l $, with $X_l = \bigoplus_{k=1}^{n_l} \cos\theta_{lk}\id_2$ and
$Y_l = \bigoplus_{k=1}^{n_l}\nu_l  \sin^2\theta_{lk}\id_2$, for $\theta_{lk}\in[0,2\pi[$.
\item $W_l$ and $Z_l$ are passive symplectic transformations acting on the system's set of modes associated with the $l$-th eigenfrequency.
\end{enumerate}
}\smallskip

Let us now unravel this statement and the restrictions it poses on the structure of GTOs, 
which will also allow us to sketch the main lines of its proof 
(whose full details are found in the SM \cite{SM}).
The transformation $S$ is just the one bringing the system Hamiltonian
into normal modes, set by the given system quadratic Hamiltonian \footnote{Whilst the environment may be set in normal form wlog, since thermal
maps do not depend on the choice of environmental basis.}.

The first step towards the statement above is realising that,
once both system and ancillas are cast into normal modes, all GTOs are obtained by letting the $n_l$ system modes pertaining to the \emph{same} eigenfrequency $\omega_l$ interact with an \emph{equal} number $n_l$ of environmental normal modes at the same frequency: $
\sig \mapsto S\left[{\rm Tr}_b \left( O\left[S^{-1}[\sig] \oplus \sig_b) \right] \right)
\right]$,
where $\sig_{b} = \oplus_{l} \nu_l \id_{2n_l}$ and $O=(\oplus_{l} O_{l})$, with 
each $O_{l}$ being a passive symplectic transformation acting on the system plus bath degenerate eigenfrequency subspace labelled by $l$ (of dimension $2n_l$). Very significantly, normal modes belonging 
to different eigenfrequency sectors do not interact during thermal operations
(this holds regardless of any correlations that may exist between the physical bath modes).

The second step to obtain the compact characterisation above is that, due to the symmetries of the problem at hand, each $O_l$ admits a very simple structure:
\be
O_l = (W_l \oplus \id_b) \circ M_l \circ (Z_l \oplus \id_b), \label{simpstruc}
\ee
where, as already stated, $W_l$ and $Z_l$ are passive symplectic on the system, and $M_l$
is a set of beam splitters independently mixing each mode $j=1,...,n_l$ with a corresponding mode of the environment:
$M_l = R^{(l)}_{n_l n_l} \oplus  \dots  \oplus R^{(l)}_{22} \oplus R^{(l)}_{11}$,
where $R^{(l)}_{kk}$ denotes a beam splitter mixing system mode $k$
(with ladder operator $\hat{a}_k=(\hat{x}_k+i\hat{p}_k)/\sqrt2$) with bath mode $k$ (with
ladder operator $\hat{b}_k$); at the Hilbert space level,
$\hat{R}^{(l)}_{kk} = {\rm e}^{(\hat{a}_k \hat{b}_k^{\dag}-\hat{a}_k^{\dag} \hat{b}_k) \theta_{lk}}$.

Thus, in a GTO, each oscillator within the degenerate frequency sector is mixed with a correspondent thermal oscillator by means of a beam splitting operation.
Tracing out the bath after such an interaction gives rise to the tensor product of phase-covariant channels that were denoted with $\Phi_l$. What is perhaps surprising is that every GTO can be simulated in this simple way, by independent interactions with the environmental modes. Besides, since the loss channels $\Phi_l$ are Markovian \cite{bucca}, the most general Gaussian thermalisation can be generated by a simple Markovian master equation. Indeed, 
GTOs are the most common as well as easiest to implement transformations, corresponding, in the normal mode basis, 
to passive optics or loss to a thermal Markovian reservoir.
\smallskip

\noindent {\em Single-mode criteria --} For each non-degenerate system eigenfrequency, a GTO
reduces to a single-mode transformation. All
single-mode passive transformations are phase shifters, 
and the transformation $Z_l$ may always be simplified by left-multiplication with phase shifters (see \cite{SM}) 
and may thus, on a single-mode, be reduced to the identity without loss of generality. Hence,
the most general GTO on a non-degenerate eigenfrequency subspace takes a very simple form indeed:\smallskip

\noindent {\bf Proposition 1 -- Single-mode GTOs.} {\em Let $\hat{H}_s = \frac12
\hat{\bf r}^{\sf T}H_s \hat{\bf r} =  \frac{\omega}{2} \hat{\bf r}^{\sf T} S S^{\sf T} \hat{\bf r}$ be a single-mode system Hamiltonian,
then the class of {GTOs} 
is given by
\be
\sig \mapsto S\left( p D_{\varphi} S^{-1} \sig S^{-1{\sf T}} D_{\varphi}^{\sf T} + (1-p) \nu_b \id_2\right)S^{\sf T} \; ,\label{1m}
\ee
with $p\in[0,1]$, $\nu_b = \frac{{\rm e}^{\beta\omega}+1}{{\rm e}^{\beta\omega}-1}$ and $D_\varphi = \left(\begin{array}{cc}
\cos\varphi & \sin\varphi \\
-\sin\varphi & \cos\varphi
\end{array}\right)$.
}\smallskip
We can now spell out the
full criterion for Gaussian state transformations through single-mode GTOs.
That is, given an input CM $\sig_i$ and an output CM $\sig_f$, is there a GTO mapping $\sig_i$ into $\sig_f$?
Here, one should recall that the most general single-mode
CM $\sig$ may be written as a rotated and squeezed thermal state:
$\sig = \nu D_{\varphi} {\rm diag}(z,1/z) D^{\sf T}_{\varphi}$, for $\varphi\in[0,2\pi[$, $z\ge 1$ and $\nu\ge1$. \smallskip

\noindent {\bf Proposition 2 -- Single-mode state transformations.} {\em Let $\hat{H}_s = \frac12
\hat{\bf r}^{\sf T}H_s \hat{\bf r} =  \frac{\omega}{2} S S^{\sf T}$ be a single-mode system Hamiltonian. An initial Gaussian state with CM $\sig_i = \nu_i S D_{\varphi_i} {\rm diag}(z_i,1/z_i) D^{\sf T}_{\varphi_i}S^{\sf T}$ may be mapped into a Gaussian state with CM $\sig_f = \nu_f SD_{\varphi_f} {\rm diag}(z_f,1/z_f) D^{\sf T}_{\varphi_f}S^{\sf T}$
via a GTO at inverse temperature $\beta$
if and only if
\be
\exists \, p\in[0,1]\, : \; \begin{array}{c}
\nu_f z_f = p \nu_i z_i + (1-p) \nu_b \\
\\
\frac{\nu_f}{z_f} = p\frac{\nu_i}{z_i} + (1-p) \nu_b \\
\end{array} \; ,
\ee
with $\nu_b = ({\rm e}^{\beta\omega}+1)/({\rm e}^{\beta\omega}-1)$. }\smallskip

Note that the parameters $\varphi_i$ and $\varphi_f$ are irrelevant to the transformation
criterion, which admits a simple geometrical representation: if one parametrises the class of single-mode
Gaussian states (in the basis of normal modes of $\hat{H}_s$ and modulo phase shifters)
in the two-dimensional space $(\nu z,\nu/z)$, one can thermally map the states $(\nu_i z_i, \nu_i/z_i)$ only into states
lying along the segment connecting $(\nu_i z_i, \nu_i/z_i)$ to $(\nu_b,\nu_b)$ (see \cite{SM}). 
Note that the squeezed states to which this criterion applies display quantum coherence
(off-diagonal elements) in the energy eigenbasis.
In the case with no squeezing, where the states have no coherence in the energy eigenbasis, the transformation criterion reduces to
$\nu_f\in[\nu_b,\nu_i]$.
In physical terms, this is equivalent to stating that GTOs send an initial thermal state 
at temperature $T_i$ into a final thermal state at temperature $T_f$ falling between $T_i$ and the environment's temperature $T$.
This complies with the thermo-majorisation and the many second laws criteria of~\cite{horodecki2013fundamental, brandao2013second} (see \cite{SM}), while the case with squeezing falls beyond the criteria's scope. Interestingly, the prediction that $T_f$ must fall between $T_i$ and $T$ differs from what happens in qubit systems and turns out to be crucial 
for the task of cooling, to which we now turn.\smallskip




\noindent {\em Algorithmic cooling -- } Let us now discuss the main repercussions of the characterisation derived above on the algorithmic cooling of  Gaussian systems.
 In the spirit of heat-bath algorithmic cooling (HBAC) \cite{park2016heat} one aims at cooling a system by alternating Gaussian unitaries and thermal operations which, 
if one allows for partial rather than complete thermalisations, 
may lead to improvements in the cooling of finite-dimensional systems \cite{rodriguez2017heat, alhambra2018heat}. 
For example, a single qubit can be cooled arbitrarily close to the ground state by applying to it Pauli $x$ unitaries interspersed with thermal operations, without the need of extra ancillas. In fact, at low enough temperatures, the required thermal operations can be approximated by resonant Jaynes-Cummings couplings to a single, initially thermal oscillator. A natural question is then if a single system oscillator can be cooled below the environment temperature in a similar fashion; that is, by unitaries on the system 
acting between the GTOs. This would be particularly advantageous because it would only require standard quadratic interaction Hamiltonians.

Here we answer this question in the negative for single-mode systems: If $\mathcal{U}_j$ are single-mode (not necessarily Gaussian) unitaries
and $\mathcal{T}_j$ arbitrary single-mode GTOs, for each $N$ the state
$
\mathcal{T}_N \circ \mathcal{U}_N \circ \dots \mathcal{T}_{1} \circ \mathcal{U}_1 [\varrho_0]
$
cannot be cooled below the minimum between the environment's entropy and the initial system entropy. 
This is the case since the output entropy of phase-covariant, single-mode Gaussian channels
at given input entropy is minimised by (Gaussian) thermal inputs
(with respect to the normal mode Hamiltonian) \cite{depalma16},
with optimal output entropy that is monotonic in the input entropy.
Thus, the best the unitaries $\mathcal{U}_j$ can do is put the state in normal form which, for given initial symplectic eigenvalue $\nu_j$, yields the output
symplectic eigenvalue $p\nu_j+(1-p)\nu_b \ge \min\{\nu_j,\nu_b\}$, so that the minimum entropy is obtained by either shielding completely from the environment or by complete thermalisation. Notice that, rather remarkably, such an entropic bound holds for any unitary operation and any input state, not necessarily Gaussian.
We also show in the SM \cite{SM} that the impossibility of lowering the system entropy below the environment's value is maintained if one extends the
class of thermal operations to include single-mode squeezed baths, which are not encountered spontaneously
in nature but may be engineered under certain controlled conditions \cite{tombesi94,lutkenhaus98,werlang08,kronwald14,klaers17}.

Cooling opportunities open up if non quadratic interaction Hamiltonians or control over the energy levels' structure are allowed, 
as is commonly assumed for quantum refrigerators~\cite{kosloff2014quantum}, or if 
some of the thermal ancillary modes can be manipulated by \emph{general} Gaussian unitaries. In point of fact, these latter schemes, unless restricted by practical constraints, allow
one to always cool any oscillator arbitrarily close to the ground state. To this aim, one may in principle
include a thermal ancillary mode at high enough frequency so that its entropy is arbitrarily low, and then swap such
a low entropy state into the system through a beam splitter acting in the unitary step
(notice that such an interaction between modes at different frequencies would not be prohibited,
since the unitary does not have to be a thermal operation in the general set-up we are considering).
This is nothing but the discrete version of sideband cooling, where excitations are extracted from the system of interest (such as a mechanical oscillator) into a coupled oscillator (such as a mode of light, in optomechanical set-ups) at higher frequency, from where they leak to the environment.

Our no-go theorem complements the impossibility
of engineering absorption refrigerators with Gaussian resources alone, pointed out in \cite{martinez13}. 
Our treatment is broader, relying on the general GTOs rather than on a specific time-evolution, and focuses on 
the system temperature, rather than heat transport between reservoirs.\smallskip

\noindent{\em Conclusions and outlook --}  We presented a full characterisation of Gaussian thermal operations, implying that they are all generated by a simple, time-local master equation, determined necessary and sufficient conditions for transformation under GTO on a single-mode and proved that no algorithmic cooling acting on a single-mode system alone can ever lower the entropy below the background or initial ones,
a fact which is relevant in practice given the broad applicability of noise models based on bilinear interactions with an
environment. 
The latter finding is intimately related to the fact that GTOs are all Markovian. As such, any dynamical trajectory reaching the thermal state must terminate there. In fact, the cooling protocol for a single qubit presented in \cite{alhambra2018heat} relied precisely on the fact that system-bath correlations can be used to cross the thermal state and achieve temperatures lower than that of the environment. This possibility is precluded, for Gaussian systems, by our no-go result. Our framework, however, sets up the scene to explore transformation conditions and more articulate cooling schemes in multimode scenarios (we refer the reader to the final section of the SM \cite{SM} for a detailed discussion of future perspectives).
\smallskip

\noindent {\em Note added --} During the completion of this article,
we became aware of closely related work \cite{gu19}, where thermal transformations are constrained to passive unitaries by design 
and several multimode necessary conditions for state transformation are discussed.\smallskip


\acknowledgments M.~Genoni played a key role by liaising between the first two authors.
Discussions with H. Jee, C. Sparaciari, G. De Palma and M. Huber, A. Levy and M. Mehboudi are also warmly acknowledged.
ML acknowledges financial support from the the European Union's Marie Sk{\l}odowska-Curie 
individual Fellowships (H2020-MSCA-IF-2017, GA794842), Spanish MINECO 
(Severo Ochoa SEV-2015-0522 and project QIBEQI FIS2016-80773-P) and 
Fundacio Cellex and Generalitat de Catalunya (CERCA Programme and SGR 875).
GA acknowledges financial support from the European Research Council under the Starting Grant GQCOP (Grant No. 637352), 
as well as from FAPESP (Grant No. 2017/07973-5).



\bibliography{Bibliography_thermodynamics2}

\widetext
\newpage

\begin{center}{\bf Supplemental Material}\end{center}

\section{First-order terms}

Since the main text considers only the purely quadratic case,
let us discuss here the effect on thermal operations of terms of the first-order in the canonical operators.
First-order terms in the bath Hamiltonian are immaterial, since they can always be set to zero by a local unitary
operation (a local phase-space displacement). They can therefore be disregarded without loss of generality, as it has been done in the paper.

Any system Hamiltonian with first-order terms, such as $\hat{H}_s = \frac12 (\hat{\bf r}-{\bf r})^{\sf T} H_s (\hat{\bf r}-{\bf r})$
can be written as $\hat{H}_s = \hat{D}^{\dag}_{\bf r} \frac12 \hat{\bf r}^{\sf T} H_s \hat{\bf r}\hat{D}_{\bf r}$, for the unitary
displacement operator $\hat{D}^{\dag}_{\bf r} = {\rm e}^{i{\bf r}^{\sf T}\Omega \hat{\bf r}}$, which
indeed just displaces the canonical operators by real quantities.
Thermal operations with respect to such a displaced Hamiltonian are therefore just given by
\be
\varrho \mapsto \hat{D}^{\dag}_{\bf r} {\mathcal T} (\hat{D}_{\bf r}\varrho\hat{D}^{\dag}_{\bf r}) \hat{D}_{\bf r} \; ,
\ee
where ${\mathcal T}$ is the thermal operation with respect to the corresponding centred Hamiltonian (with no first-order term),
as derived in the main text. Clearly, the displacement does not generally commute with ${\mathcal T}$, so that the net effect of a thermal
operation will involve a finite displacement of the first moments (which would be very easy to evaluate in specific cases).

In the main text, we also stated without proof that no Hamiltonian with strictly positive Hamiltonian matrix commute with displacement operators: this is immediately apparent since any translation of $\hat{\bf r}$ in $\hat{\bf r}^{\sf T} H \hat{\bf r}$ will always produce a nonzero
shift to the value of the operator if $H>0$. This would not be the case for a semi-definite Hamiltonian, such as the free
Hamiltonian $\hat{p}^2$, which is obviously invariant under translations of the $\hat{x}$ operator.

\section{Symplectic rendition of thermal operations}

We can work in the local system and bath symplectic bases where the local Hamiltonian
matrices are in normal form, and then consider the most general interaction Hamiltonian matrix $H_I$.
For the bath, this can be done without loss of generality, since it just corresponds to a choice of basis of a subsystem which will be ultimately
traced out. For the system, such an assumption will be relaxed by including the action of the symplectic $S$ that brings
the system Hamiltonian to normal modes.

In such bases, one has $H_s = \bigoplus_{l} \omega_{s,l} \id_{2n_l}$ and
$H_b = \bigoplus_{l} \omega_{b,l} \id_{2m_l}$. Bear in mind that, because we allow for the addition of ancillary modes with arbitrary Hamiltonians, the bath eigenfrequencies $\omega_{b,l}$ and degeneracies $m_l$ are whatever
we like them to be. In other words, the only input parameters determining the set of thermal operations are the
system's eigenfrequencies $\omega_{s,l}$ and degeneracies $n_l$, as well as the inverse temperature $\beta$.

Notice now that a necessary condition for the Hamiltonian operator $\hat{H}_I$ to commute with $\hat{H}_s+\hat{H}_b$ is that
the unitary transformations generated by exponentiating $i\hat{H}_I$ leave $\hat{H}_s+\hat{H}_b$ unchanged. In terms of quadratic Hamiltonians,
this is equivalent to stating that the symplectic transformations ${\rm e}^{-\Omega H_I t}$ \cite{bucca}
must belong to the subgroup of transformations
that leave $H_s\oplus H_b = \bigoplus_l \omega_{s,l} \id_{2n_l} \bigoplus_{l} \omega_{b,l} \id_{2m_l}$ unchanged when acting by congruence.
But such an isotropy group is easily characterised:\medskip

\noindent{\bf Lemma 1 -- Isotropy group of normal form matrices.}
{\em The symplectic isotropy group of the transformation $Y= \bigoplus_{l} \omega_l \id_{2 d_l}$ is given by the
direct sum of the compact symplectic subgroups $K(2d_l) = Sp_{2d_l,{\mathbbm R}} \cap SO(2d_l)$, each acting on
the $2d_l$-dimensional subspace pertaining to a certain eigenfrequency $\omega_l$.}\smallskip

\noindent {\em Proof.} Let $K$ be a symplectic transformation part of the isotropy group. Then, by definition $KYK^{\sf T} = Y$
and $K\Omega K^{\sf T} = \Omega$. Recalling that $K$ is invertible, it is easy to show that the previous two equations imply
$[K,Y\Omega]=0$. If $K$ is written in terms of $2\times 2$ sub-blocks $K_{jk}$, as per
\be
K = \left(\begin{array}{ccc}
K_{11} &\cdots& K_{1d} \\
\vdots & \ddots & \vdots \\
K_{d1} & \cdots & K_{dd}
\end{array}\right) ,
\ee
then the simple form of $Y\Omega$ allows one to reduce the commutation condition with $K$ to a condition on the
sub-blocks:
\be
\omega_k  K_{jk} \Omega_1 - \omega_j \Omega_1 K_{jk}  = 0  \;
\ee
(where $\Omega_1$ is the $2\times 2$ symplectic form on a single mode).
Writing
\be
K_{jk} = \left(\begin{array}{cc}
a & b \\
c & d
\end{array}\right) \; ,
\ee
this yields the set of equations
\be
(a+d)(\omega_k-\omega_j) = (c-b)(\omega_j-\omega_k) = (a-d)(\omega_j+\omega_k) = (b+c)(\omega_j+\omega_k) = 0 \;
\ee
which, for $\omega_j\neq \omega_k$, imply $K_{jk}=0$. Therefore, the isotropy transformation $K$ must
be block-diagonal with respect to subspaces associated with distinct symplectic eigenvalues of $Y$,
and must be a direct sum of symplectic orthogonal transformations on each such subspace
(since any such transformation clearly preserves $Y$). \hspace{\stretch{1}} $\square$ \smallskip

Let us remark that one may show that all of these isotropy
transformations are generated by Hamiltonians that commute with the Hamiltonian they preserve,
so that each of them does indeed define a legitimate Gaussian thermal operation.
The orthogonal symplectic transformations that form the isotropy subgroups
are also referred to as ``passive'' in the quantum optics tradition, since they preserve the
number of photons.

By virtue of the statement above, Gaussian thermal operations act separately on each of the system's phase space
sectors pertaining to a different eigenfrequency. Besides the passive, symplectic transformations acting on such subspaces,
which are obviously all thermal, less trivial examples of Gaussian thermal operations are obtained by appending
to each degenerate subspace with eigenfrequency $\omega_l$ a set of bath modes at the very same frequency $\omega_l$.
Such modes are all prepared, before the unitary transformation, in the (Gaussian) thermal Gibbs state with covariance matrix
$\nu_l \id_2$, with $\nu_l = ({\rm e}^{\beta \omega_l}+1)/({\rm e}^{\beta \omega_l}-1)$, and we can add as many as we like
(see Ref.~\cite{bucca} for the formula relating frequency and temperature to the symplectic eigenvalue).

In order to complete our characterisation of Gaussian thermal operations, we now set out to characterise the set of Gaussian CP-maps
obtained by letting an input Gaussian state of $n$ modes, with arbitrary covariance matrix $\sig$,
interact with an environment with covariance matrix $\nu \id_{2m}$, through a global passive symplectic transformation, for all integer $m$.

\section{Unitary representation of the compact subgroup}

It is well known that, by adopting a representation in terms of annihilation and creation operators, passive symplectic transformations
in dimension $2d$
may be represented as $\left(\begin{array}{cc} U & 0 \\ 0 & U^* \end{array}\right)$, where $U\in U(d)$
(in the field theory tradition, this is known as the `Bogoliubov' representation of passive symplectic operations) \cite{bucca}.
Such an isomorphism
between $K(2d)$ and $U(d)$ will be very advantageous in describing arbitrary passive symplectic acting on the degenerate
eigenfrequency sectors of the system plus bath Hamiltonian.

In this notation, which, in each eigenfrequency sector, corresponds to taking the basis of operators 
$$ (\hat{a}^s_1,\ldots,\hat{a}^s_{n_l},\hat{a}^b_1,\ldots,\hat{a}^b_{m_l},\hat{a}^{\dag s}_1,\ldots,\hat{a}^{\dag s}_{n_l},\hat{a}^{\dag b}_1,\ldots,\hat{a}^{\dag b}_{m_l})^{\sf T}$$ 
(with $s$ denoting the system and $b$ the bath), the global, initial CM describing system and bath may be written as
$$
\left(\begin{array}{cccc} \sig_{a_l a_l^{\dag}} & 0 &\sig_{a_l a_l} & 0 \\
0 & \nu_l \id_{m_l} & 0 & 0 \\
 \sig_{a_l a_l}^{\dag} & 0 &\sig_{a_l a_l^{\dag}} & 0 \\
 0&0&0&\nu_l\id_{m_l}\end{array} \right) \; ,
$$
where $\sig_{a_l a_l^{\dag}}$ is an $n_l\times n_l$ hermitian matrix reporting the values of the symmetrised covariances of all pairs of
system annihilation and creation operators (one each), whilst $\sig_{a a}$ contains the covariances of pairs of annihilation operators.
The blocks $\nu_l \id_{m_l}$ correspond to the covariances of the initial thermal state of the $m_l$ bath modes.

Inspection of the initial CM above reveals that the CP-map obtained by letting such an initial state evolve through a global
passive represented by $U$ is invariant under right multiplication of $U$ by an arbitrary bath unitary $Z_{m_l}$.
Besides, one can also left-multiply $U$ by another, generally different, bath unitary $W_{m_l}$, since the bath is ultimately traced out
(corresponding, in the CM formalism, to pinching the relevant part of the matrix).
These symmetries will be key to what follows.

Note also that, under such a choice of basis, `standard' beam splitters may be represented as real two-dimensional rotations,
which we shall denote with the letter $R$ below: $R=\left(\begin{array}{cc}
\cos\theta & \sin\theta \\
-\sin\theta & \cos\theta
\end{array}\right)$ for $\theta\in[0,2\pi[$ (acting on the relevant components, which will be specified through indexes below).

\section{Simplifying the unitary matrix}

As we just saw, a global $(n_l+m_l)$-dimensional unitary $U$, that determines the thermal Gaussian CP-map
by acting on system and bath in a certain
eigenfrequency sector, may be simplified by acting on the left and right through a local, bath unitary, as in the lemma below.\medskip

\noindent {\bf Lemma 2 -- Triangularisation of off-diagonal blocks.}
{\em Let $U$ be an $(n+m)\times (n+m)$ matrix with $m\ge n$. Two $m\times m$ unitary matrices $U_{m}$ and $V_m$
always exist such that:
\begin{equation}
({\id_n}\oplus U_m) U ({\id_n}\oplus V_m) = \left(\begin{array}{cc} \alpha & \beta \\
\gamma^{\sf T} & \delta
\end{array}\right) \; ,
\end{equation}
with
\be\hspace*{-2cm}
\beta = \left(\begin{array}{cccccccc}b_{11} & 0 &\cdots &\cdots & \cdots & \cdots & \cdots& 0 \\
b_{21} & b_{22} & 0 & \cdots &\cdots&\cdots &\cdots& 0 \\
b_{31} & b_{32} & b_{33} & 0 & \cdots&\cdots & \cdots&0 \\
\vdots &&&\ddots&\ddots&&&\vdots \\
b_{n1} & \cdots&\cdots&\cdots & b_{nn} & 0 &\cdots &0
\end{array}\right) \; , \quad
\gamma = \left(\begin{array}{cccccccc}g_{11} & 0 &\cdots &\cdots & \cdots & \cdots & \cdots& 0 \\
g_{21} & g_{22} & 0 & \cdots &\cdots&\cdots &\cdots& 0 \\
g_{31} & g_{32} & g_{33} & 0 & \cdots&\cdots & \cdots&0 \\
\vdots &&&\ddots&\ddots&&&\vdots \\
g_{n1} & \cdots&\cdots&\cdots & g_{nn} & 0 &\cdots &0
\end{array}\right) \label{offb}
\ee
}\smallskip

\noindent {\em Proof.} {Any $n$ $m$-dimensional vectors (such as the row vectors of the initial form of the block $\beta$)
can be put into the form above by a $m$-dimensional unitary operation $V_m$.
Likewise for $\gamma$, acting with $U_m$ from the other side. This is equivalent to the QR decomposition, whereby
any matrix can be put in triangular form by acting from a side with a unitary matrix.} \hspace{\stretch{1}}$\square$\smallskip

Because of the previous lemma we can restrict, without loss of generality, to baths with the same number of modes as the system ($n=m$) \footnote{The general possibility
of reproducing Gaussian CP-maps by doubling the number of modes in the system is well known. However, this explicit argument sorts out any complication that might arise from
the specific restrictions of the problem in hand.}.
Now, a $U$ with off-diagonal blocks of the form above can be decomposed as follows.\medskip

\noindent {\bf Lemma 3 -- Cosine-sine decomposition.} {\em Let $U$ be a $2n\times 2n$ unitary matrix, then
\be
U = (W \oplus X) (R_{nn}\oplus \ldots \oplus R_{11}) (Z \oplus Y) \, , \label{artdeco}
\ee
where $W$, $X$, $Z$ and $Y$ are $n\times n$ unitary matrices, while
$R_{jj}$ is a (real) beam splitter between the $j$-th and the $(n+j)$-th mode.}\medskip

This is a standard decomposition of unitary matrices, which follows from taking the singular value decomposition of the two $n\times n$
off-diagonal blocks through the local unitaries and then apply the unitarity conditions (see, e.g., \cite{fuhra18}).

Let us notice, en passant, that minor variations of the lemma above may be employed to obtain an explicit proof
of the well known results that (i) any unitary may be decomposed into two-level unitaries, and (ii) any passive symplectic transformation is the product of beam splitters and phase shifters.

Lemma 3 is incredibly revealing to the purpose of simplifying Gaussian thermal operations: indeed
it is telling us that, in each eigenfrequency subsector,
and up to an initial and final passive symplectic acting on the system ($Z$ and $W$, respectively), the action of a thermal map
boils down to mixing each system normal mode with a bath mode, independently, through a standard beam splitter.
The local unitary transformations on the bath $X$ and $Y$ can be completely disregarded: the former because it acts at the very end,
just before the bath is traced out, the latter because the initial bath state, given by a thermal state on modes with degenerate
normal frequency and hence with CM proportional to the identity, is invariant under passive transformations.

Note also that the decomposition above is slightly redundant, as it involves $4n^2+n$ real degrees of freedom ($n^2$ per unitary, plus
$n$ for the $n$ mixing angles of the beam splitters), against the $4n^2$ degrees of freedom of a $2n$-dimensional unitary.
In fact, one of the four unitaries is not completely arbitrary, but can be
simplified by multiplication on a side by any diagonal matrix of complex phases (corresponding to a tensor product of
single-mode phase shifters in physical set-ups). It is easy to see that such a multiplication may be absorbed by redefining the other
unitaries without affecting the singular values of the off-diagonal blocks (which, effectively, set the beam splitters's angles).
To our purpose, it will be convenient to simplify the matrix $Z$, although $W$ might also have been chosen.

\section{Parametrisation of general Gaussian thermal operations}

All the above was derived for the system normal modes, whose local Hamiltonian matrix we shall denote hereafter with $\omega_s$.
The most general local Hamiltonian matrix is therefore $S \omega_s  S^{\sf T}$,
where $S$ is any local symplectic transformation on the system.

\begin{figure}[t!]
\captionsetup{justification=justified}\subfloat[\label{left}]{\includegraphics[valign=c]{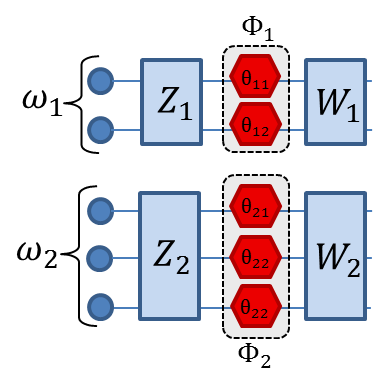}}
\hspace*{1cm}
\subfloat[\label{right}]{
\includegraphics[valign=c]{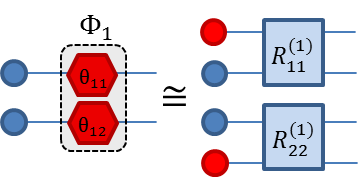} \vphantom{\includegraphics[valign=c]{decomposition2.png}}%
}
\begin{flushleft}
\caption{Schematics of a GTO acting in the normal-mode basis:
(a) a 5-mode system, with degenerate eigenfrequencies $\omega_1$, pertaining to two modes,
and $\omega_2$, pertaining to three modes, undergoes the initial passive symplectics $Z_1$ and $Z_2$, followed by a tensor
product of phase-covariant channels $\Phi_1$ and $\Phi_2$; in turn, each $\Phi_j$ is the tensor product of
phase-covariant channels $\theta_{jk}$, each acting on a mode separately; finally, the passive symplectics $W_1$
and $W_2$ act separately on the degenerate eigenspaces; (b) each phase-covariant channel $\theta_{jk}$
is shown to result from the mixing of the system mode at a beam splitter, whose transmittivity sets the parameter
$\theta_{jk}$ (here, for simplicity, the parameter $\theta_{jk}$ also denotes the single-mode channel itself).
\label{Algo}}
\end{flushleft}
\end{figure}

Above, we determined and simplified {\em all} of the global symplectic transformations $S_{I}$ that preserve $\omega_S\oplus H_b$,
where $H_b$ is the bath Hamiltonian matrix.
It follows that the whole set of global symplectic that preserves a general quadratic Hamiltonian not in normal form, as given above, is just
\be
S_{I}' = (S\oplus \id_b) S_{I} (S^{-1}\oplus \id_b) \; .
\ee
This fact, along with the decomposition (\ref{artdeco}) and the basic piece of knowledge that
a beam splitting interaction with an environmental mode gives rise to the phase-covariant
CP-map $\Phi$, that maps a single-mode CM $\sig$ according to
$\Phi(\sig) = \cos^2\theta \sig + \sin^2\theta \nu_b \id_2$, leads directly
to the general characterisation of Gaussian thermal operations given by Theorem 1, which is illustrated in Fig.~\ref{Algo}.

As explained in the previous section, there is some residual freedom in the constructive characterisation of Theorem 1.
Because of the residual ambiguity in the cosine-sine decomposition,
whilst the operations $W_l$ may be taken as completely arbitrary passive symplectic transformations, the transformations $Z_l$ are passive symplectic operations that can be simplified by the action of a
tensor product of phase shifters acting on them from the left: each of them thus bear $n_l^2-n_l$
free parameters (recalling that $n_l^2$ is the number of parameters in an arbitrary passive symplectic transformation).
Therefore, up to the transformation $S$, a GTO acting on a degenerate eigenfrequency sector comprising $n_l$ modes is parametrised by $2n_l^2+1$ parameters (one of them being the inverse temperature $\beta$).

\subsection{The single-mode case}

For $n_l=1$, which covers all systems with non-degenerate eigenfrequencies,
the only local passive transformation is the phase shifter 
$D_{\varphi}$ given, in the $(\hat{x},\hat{p})$
basis, by
\be
D_{\varphi} = \left(\begin{array}{cc}\cos\varphi & \sin\varphi \\
-\sin\varphi & \cos\varphi
\end{array}\right) \; .
\ee
As discussed above, the passive transformation $Z_l$ entering Eq.~(2) of the main text
can be simplified through left-multiplication by a phase shifter, and may thus be reduced to the identity without loss of generality
in the single-mode case.
Setting $p=\cos^2\theta$,
one is therefore left with the expression reported in Proposition 1 and Eq.~(3) of the main text.

\section{Single-mode state transformations}

Let us restate the most general thermal mapping
for a non-degenerate (single-mode) system frequency:
\be
\sig_{f} = p (S D_{\varphi} S^{-1} \sig_{i} S^{-1{\sf T}} D_{\varphi}^{\sf T} S^{\sf T}) + (1-p) \nu_b {S}{S}^{\sf T} \; ,
\ee
which has been written in terms of the initial and final CMs $\sig_{i}$ and $\sig_{f}$
in view of our next objective, which is characterising allowed
mappings between pairs of states at given $\nu_b$ (temperature).

Clearly, one can re-write the initial and final CMs in the normal basis to obtain a condition independent from $S$.
Formally, one
can act on the left and right hand sides with $S$ and obtain
a condition for the transformed input and output $\sig_{i,f}'=S^{-1}\sig_{i,f} S^{-1 \sf T}$:
\be
\sig_{f}' = p ( D_{\varphi}  \sig_{i}'  D_{\varphi}^{\sf T} ) + (1-p) \nu_b \id_2  \; . \label{map}
\ee

Single-mode Gaussian states are particularly simple,
as can be seen by applying the symplectic singular value decomposition to the Williamson form
of a state \cite{bucca}. Their most general form is $\sig'_{i,f} = \nu_{i,f} D_{i,f} Z_{i,f} D_{i,f}^{\sf T}$,
where $\nu_{i,f}$ are the initial and final symplectic eigenvalues
(which determines {\em any} entropy in the single-mode case), $D_{i,f}$ are single-mode rotations and $Z_{i,f}={\rm diag}(z_{i,f},z_{i,f}^{-1})$,
and we can assume $z_{i,f}\ge 1$ without loss of generality (since phase space rotations allow one to invert $z_{i,f}$).

Since thermal mappings are rotationally invariant in phase space, one can always match the optical phases of input and output, and we can therefore disregard the rotations altogether.
One is then left with the following necessary and sufficient conditions for state transformations:
\be
\exists\; p\in[0,1] \; : \; \left\{ \begin{array}{ccc}
z_f\nu_f & = & pz_i \nu_i + (1-p) \nu_b  \, , \\
&&\\
\frac{\nu_f}{z_f} & = & p\frac{\nu_i}{z_i} + (1-p) \nu_b \, .
\end{array} \right. \label{syst}
\ee

\subsubsection{Isotropic states}

In the absence of squeezing ($z_{i,f}=1$),
the situation is very simple to depict, as the conditions above lead to the necessary and sufficient condition that $\nu_{f}$
must lie between $\nu_b$ and $\nu_i$.

Note that, for single-mode Gaussian states, the free energy $F$ in the normal mode basis (at eigenfrequency $\omega$)
may be easily expressed as
(see \cite{bucca} for a formula expressing the von Neumann entropy of a Gaussian state as a function
of the symplectic eigenvalue $\nu_b$)
\be
F = \frac14 \omega \nu_b (z+\frac1z) - \frac{1}{\beta} \left[ \frac{\nu_b+1}{2}\ln\left(\frac{\nu_b+1}{2}\right)
-\frac{\nu_b-1}{2}\ln\left(\frac{\nu_b-1}{2}\right) \right] \; . \ \label{freef}
\ee
For $z=1$ and at given $\beta$, such a function of $\nu_b$ has a single minimum at the environmental value
$\nu_b = \frac{{\rm e}^{\beta \omega+1}}{{\rm e}^{\beta \omega}-1}$.
Therefore, the transformation criterion $\nu_f\in[\nu_b,\nu_i]$ (regardless of the ordering of $\nu_b$ and $\nu_i$)
tells us that, even in the absence of squeezing, the decrease in the free energy is necessary (as it always is, since thermal operations have thermal fixed points)
but not sufficient for two Gaussian states to be thermally connectable through an environment at inverse temperature $\beta$.

Notice also that, since such states are diagonal in the energy eigenbasis, the hierarchy
of free energy criteria pointed out in \cite{brandao2013second}
will apply to them. However, under the additional assumptions of a single-mode system in a Gaussian state, all such 
thermal transformation criteria coalesce to a single one, since all Renyi entropies are determined by a single quantity.

\subsubsection{General squeezed states}

Solving the system above for $p$ yields
\be
p = \frac{z_f\nu_f-\nu_b}{z_i\nu_i-\nu_b}=\frac{\nu_f/z_f-\nu_b}{\nu_i/z_i-\nu_b} \; ,
\ee
whose boundedness ($0\le p\le 1$) gives the necessary and sufficient conditions for possible transformations.

\begin{figure}[t!]
\includegraphics[scale=0.8]{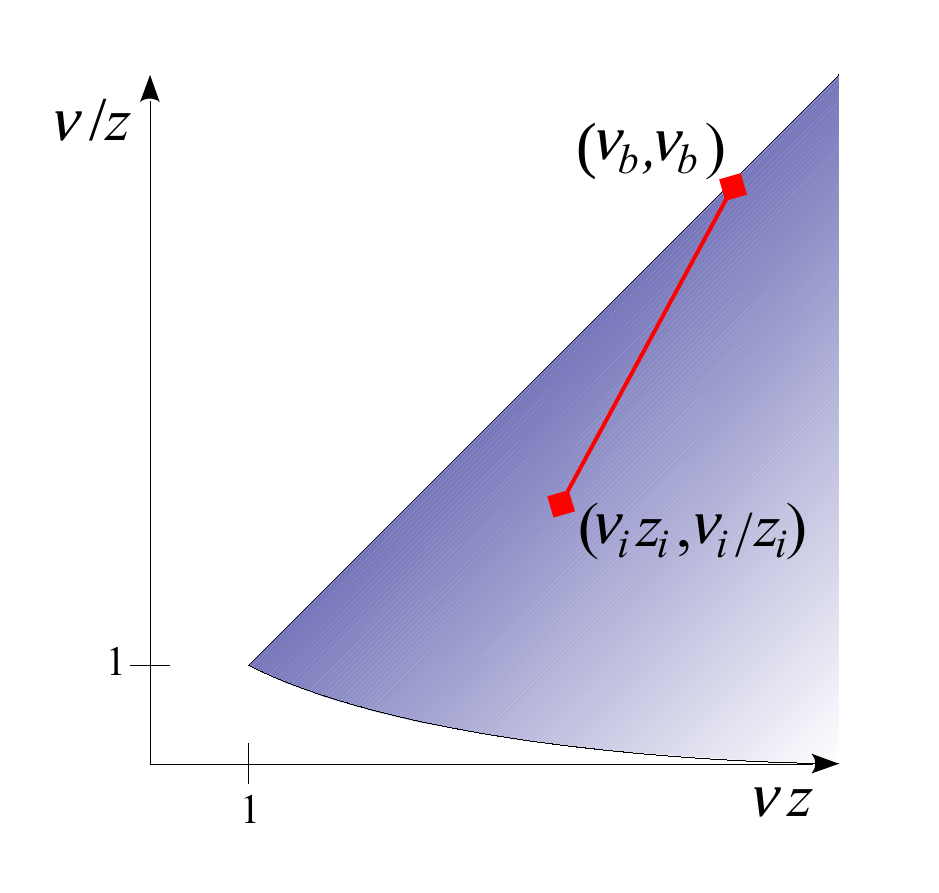}
\caption{Transformation criterion for single-mode systems. The shaded area contains all single-mode Gaussian states
which, up to rotations and first moments,
are parametrised by the symplectic eigenvalue $\nu\ge 1$ and squeezing parameter $z\ge 1$ (in the normal-mode basis
of the system Hamiltonian).
Non-squeezed states (which, for zero first-moments, are diagonal in the energy eigenbasis) lie on the $z=1$ line that bisects
the two axes. Given an environmental symplectic eigenvalue $\nu_b$ (set by frequency and temperature),
one has that an initial state parameterised by $\nu_i$ and $z_i$ may be mapped into a final state with $\nu_f$ and $z_f$
if and only if the point $(\nu_f z_f,\nu_f/z_f)$ belongs to the segment connecting $(\nu_i z_i,\nu_i/z_i)$ to $(\nu_b,\nu_b)$
(represented in red between diamonds on the graph). \label{Transf}}
\end{figure}

Direct inspection of (\ref{syst}) reveals the whole geometric nature of such a necessary and sufficient condition,
illustrated in Fig.~\ref{Transf}.
Given $\nu_b$, as well as the input $\nu_i$ and $z_i$, it is convenient to parametrise the possible output state
in the space $\nu_f z_f$ and $\nu_f/z_f$, for $z_f\ge 1$.
For a thermal mapping to be possible, it is necessary that such variables belong to the interval $[\nu_b,\nu_i z_i]$
and $[\nu_b,\nu_i /z_i]$ (denoting, up to the proper ordering, the interval between the two values).
The necessary and sufficient condition is that $(\nu_f z_f,\nu_f/z_f)$ belong to
the diagonal of such an interval, joining $(\nu_b,\nu_b)$ to $(\nu_i z_i,\nu_i/z_i)$. Notice that for $z_i=1$
the interval becomes a square and the conditions reduce to $z_f=1$ and $\nu_f\in[\nu_b,\nu_i]$.
The effect of the initial squeezing is precisely to make such a square oblong.

Simple necessary conditions about $\nu_f$ and $z_f$ may also be obtained as follows.
Taking the ratio of the two equations in (\ref{syst}), one gets
\be
z_f^2 = \frac{\left(z_i^2\frac{\nu_i}{\nu_b}-z_i\right)p +z_i}{\left(\frac{\nu_i}{\nu_b}-z_i\right)p+z_i} \le z_i^2 \; ,
\ee
which can be shown by observing that the derivative of the function above
with respect to $p$ is the always positive $\frac{(z_i^2-1)z_i \nu_b \nu_i}{\left[(1-p)z_i\nu_b+p\nu_i\right]^2}$
(recalling that $z_i\ge 1$).
Clearly, mixing with a non-squeezed state cannot increase the squeezing.

The product of the two equations in (\ref{syst}) instead yields
\be
\nu_f^2 = p^2 \nu_i^2 + (1-p)^2 \nu_b^2 + p(1-p) \nu_b \nu_i (z_i+z_i^{-1}) \ge \left( \nu_i p + \nu_b (1-p) \right)^2 \; .
\ee
Since $\nu_f$ is larger than a convex combination of $\nu_i$ and $\nu_b$,
it must also be
\be
\nu_f \ge \min (\nu_i,\nu_b) \; .
\ee
However, at variance with the isotropic case, it can be that $\nu_f \ge \nu_i \ge \nu_b$: for instance,
for $\nu=\nu_i=2$ and $z_i=4$, one has the allowed thermal transformation into $\nu_f=5/2$ and
$z_f=2$. Rather interestingly, thermal operations can turn squeezing into thermal energy.

The explicit proofs above, which we worked out within the Gaussian framework,
are subsumed by the general Hilbert space statements
that were recently derived to settle the question of the minimum output entropy of single-mode
phase-covariant Gaussian channels, which comprise all GTOs. This line of enquiry has shown that
the minimum output purity of such channels is obtained for a coherent state input
(any Gaussian state with CM $\id_2$) \cite{mari14,giovannetti15},
and that the minimum output purity at given input entropy is attained by a thermal input (any Gaussian state with CM
$\nu_i\id_2$, with $\nu_i$ set by the input entropy) \cite{depalma16}.
Applying such results yield a minimum output symplectic eigenvalue
equal to $p\nu_i + (1-p)\nu_b \ge \min\{\nu_i,\nu\}$.


\section{Squeezed baths}

Given the current popularity of reservoir engineering approaches, it is interesting to extend our treatment to the case
where the Gaussian state of the single-mode bath interacting with the system can be made completely arbitrary.
This would encompass all instances of squeezed baths \cite{lutkenhaus98,werlang08,kronwald14,klaers17}.
In the normal mode basis, it would correspond to the following mapping:
\be
\sig_{f}' = p ( D_{\varphi}  \sig_{i}'  D_{\varphi}^{\sf T} ) + (1-p) \nu_b \sig_b \; , \label{sqmap}
\ee
where $\sig_b$ is any covariance matrix of a pure Gaussian state (the finite entropy of the environment is accounted for by $\nu_b$,
as above); that is, $\sig_b$ is any symmetric matrix with determinant $1$.

We intend to work out conditions for state transformations under the extended thermal mapping of Eq.~(\ref{sqmap}).

Notice that, due to the presence of $D_{\varphi}$ and to the complete freedom in choosing $\sig_b$, an arbitrary
rotation may be applied on $\sig_{f}'$. We can therefore assume a diagonal $\sig_f' = \nu_f Z_{f}$,
and a general $\sig_i' = \nu_i D_{\vartheta} Z_i D_{\vartheta}^{\sf T}$.

Since $\sig_b$ is any symmetric matrix with determinant $1$, one has that
given $\nu_b$, in order for a thermal transition from $\sig_i'$ to $\sig_f'$ to be possible, there must exist
a $p\in[0,1]$ such that the matrix
$$
\sig_f' - p \sig_i'
$$
has determinant $\nu_b^2 (1-p)^2$. In order to obtain a necessary and sufficient condition, one has to
also make sure that the matrix above is positive semi-definite (a condition which the determinant alone cannot probe) \footnote{Although a CM must be
strictly positive, imposing positive semi-definiteness is sufficient in the two-dimensional, single-mode case,
under the added prescription of a positive determinant. This will also yield the correct conditions for $p=1$, when the determinant is actually zero.}.

The determinant of a sum of $2\times 2$ matrices can be expressed through the well known formula:
\be
{\rm Det}\left[\sig_f' - p \sig_i' \right] =
{\rm Det}\left[\sig_f'\right] + p^2 {\rm Det}\left[\sig_i' \right] - p {\rm Det}\left[\sig_f'\right]{\rm Tr}\left[\sig_f'^{-1}\sig_i' \right] \; .
\ee
In terms of the parameters introduced above that determine $\sig_i'$ and $\sig_f'$, one has
${\rm Det}\left[\sig_f'\right] = \nu_f^2$, ${\rm Det}\left[\sig_i'\right]=\nu_i^2$ and
${\rm Det}\left[\sig_f'\right] {\rm Tr}\left[\sig_f'^{-1}\sig_i' \right] = 2 \xi \nu_i \nu_f$,
with
\be
\xi = \frac12 \left[ \cos^2\vartheta \left(\frac{z_i}{z_f}+\frac{z_f}{z_i}\right)
+ \sin^2\vartheta \left( z_i z_f + \frac{1}{z_i z_f}\right)  \right] \; ,
\ee
so that one obtains the necessary condition for thermal mapping:
\be
\nu_f^2 + p^2 \nu_i^2 - 2 p \xi \nu_i\nu_f = (1-p)^2 \nu_b^2 \; .
\ee
If the above is satisfied, sufficiency is established by ensuring that any one-dimensional pinching of the matrix
$\sig_f' - p \sig_i' $ is positive semi-definite, which leads to the following set of necessary and sufficient conditions
for thermal mappings (including the possibility of squeezed baths):
\be
\exists \; p \in[0,1] \; : \; \left\{ \begin{array}{l}
\nu_f^2 + p^2 \nu_i^2 - 2 p \xi \nu_i\nu_f = (1-p)^2 \nu_b^2 \, , \\
\\
z_f \nu_f- p\nu_i \left(\cos^2 \vartheta z_i + \sin^2\vartheta z_{i}^{-1}\right) \ge 0 \, .
\end{array} \right. \label{systg}
\ee
Note that, here, $\vartheta$ is the difference in optical phase between the input and the output state,
a relevant physical parameter in non phase-covariant transformations.
These conditions can be systematically checked, by verifying that one of the
solutions
\be
p = \frac{\nu_b^2-\xi \nu_i \nu_f \mp \sqrt{(\nu_b^2-\xi \nu_i\nu_f)^2 - (\nu_b^2-\nu_i^2)(\nu_b^2-\nu_f^2)}}{\nu_b^2-\nu_i^2}
\ee
is between $0$ and $1$, and that the second of (\ref{systg}) is satisfied.

Moreover, since $\xi\ge 1$, one has
$(1-p)^2 \nu_b^2 \le (\nu_f - p \nu_i)^2$ which -- together with the inequality $\nu_f^2\ge p^2\nu_i^2$ from Eq.~(\ref{sqmap}) [recall that ${\rm Det}{(A+B)}
\ge {\rm Det}A +{\rm Det}B$ if $A$ and $B$ are positive] -- leads to the general inequality
\be
\nu_f \ge p\nu_i + (1-p) \nu_b \; ,
\ee
whence
\be
\nu_f \ge \min(\nu_i,\nu_b)
\ee
that, as mentioned in the main text, is thus shown to hold for completely arbitrary Gaussian baths.

{\section{Comparison with thermo-majorisation}}

In the case of a single bosonic mode with energy spacing $E$, all Gaussian states with no squeezing
and zero first moments can be parametrized by an inverse temperature $\beta_i$ through the occupation probability $p_n = (1- e^{-\beta_i E})e^{-\beta_i E n}$. Similarly, the Gaussian final states will be described by $q_n = (1- e^{-\beta_f E})e^{-\beta_f E n}$. The thermal distribution is $g_n = (1- e^{-\beta E})e^{-\beta E n}$.

Thermo-majorisation is a prescription for computing which final states can be achieved under thermal operations when initial and final states are diagonal in the energy basis, as in this case. The prescription is as follows: first construct the thermo-majorisation curve of the initial and final states. The thermo-majorisation curve of the initial state is obtained by sorting the probabilities $p_n$ by a permutation $\pi_i$ such that
\begin{equation}
\frac{p_{\pi_i(0)}}{g_{\pi_i(0)}} \geq \frac{p_{\pi_i(1)}}{g_{\pi_i(1)}} \geq \frac{p_{\pi_i(2)}}{g_{\pi_i(2)}} \geq \dots
\end{equation}
and then forming the piecewise linear curve in $\mathbb{R}^2$ obtained by joining the points $(0,0)$ and $\{(\sum_{j=0}^{n}g_{\pi_i(j)}, 
\sum_{j=0}^{n} p_{\pi_i(j)} )\}$ for $n=0,1,\dots, \infty$. Similarly, we find a permutation $\pi_f$ for $q_n$ and construct the thermo-majorisation curve of the final state. Then a thermal operation exists mapping $p_n$ into $q_n$ if and only if the thermo-majorisation curve of $p_n$ lies all above that of $q_n$. A word of warning: the thermo-majorisation criterion has been rigorously proved for arbitrary finite dimensional systems. The present case, involving a harmonic oscillator, should hence be treated with care, e.g. by definition of appropriate cutoffs. Here we will content ourselves with sketching an argument, which can easily be made rigorous by introducing 
arbitrary high cutoffs, showing that our condition for state transformations complies with thermo-majorisation in the absence of squeezing.

Let us now show that $\beta_f$ cannot lie outside the interval between $\beta_i$ and $\beta$. Suppose $\beta_i < \beta < \beta_f$ (the impossibility of the opposite case, $\beta_f < \beta < \beta_i$, will trivially follow). Then $\pi_i$ sorts $n$ from $+\infty$ down to $0$, whereas $\pi_f$ sorts $n$ from $0$ to $+\infty$. Since $\frac{p_{\pi_i(n)}}{g_{\pi_i(n)}}\propto e^{(\beta - \beta_i) E n}  \left(\frac{q_{\pi_f(n)}}{g_{\pi_f(n)}} \propto e^{(\beta - \beta_f) E n}\right)$ is the slope of the $n$-th segment of the thermo-majorisation curve of $p_n$ ($q_n$), we reach the following conclusions: \medskip

\begin{enumerate}
\item The slope of the thermo-majorisation curve of $p_n$ is $\infty$ at $n=0$ and non-zero as $n \rightarrow \infty$;
\item  The slope of the thermo-majorisation curve of $q_n$ is finite at $n=0$ and $0$ as $n \rightarrow \infty$.
\end{enumerate}

These two facts imply that the two thermo-majorisation curves intersect. Hence there is neither a thermal operation mapping the initial state $\beta_i$ into $\beta_f$, nor is there a thermal operation mapping $\beta_f$ into $\beta_i$. This implies that one must have $\beta_f$ in the interval between $\beta$ and $\beta_i$, as claimed.

\subsection{Thermo-majorisation and second laws: some open questions}

Consistency with earlier results \cite{horodecki2013fundamental} requires that our single-mode condition implies thermo-majorization. However, the proof sketched above shows that the converse also holds (thermomajorisation $\Rightarrow$ single-mode condition), and this is nontrivial. The reason this equivalence is nontrivial is that thermo-majorisation is equivalent to the existence of a thermal environment, and an energy preserving unitary coupling system and environment, such that the initial state of the system, $p_n$, is mapped into $q_n$. Our single mode condition, on other other hand, additionally ensures that environment and unitary are associated to quadratic Hamiltonians. For a single mode, however, the weaker conditon (thermo-majorisation) is equivalent to the stronger (single-mode criterion). This leads to conjecture this situation extends beyond single-mode scenario, to general energy diagonal Gaussian systems. In other words, it might be that thermo-majorization, when applied to Gaussian distributions, is equivalent to the existence of a GTO mapping between them. This is the first open question we leave here.

Furthermore, we discuss the connection with the work \cite{brandao2013second}, where a set of constraints collectively known under the name of ``many second laws'' were analysed. Brushing some details aside, these are necessary and sufficient conditions for the existence of a transformation between two energy diagonal, finite-dimensional states, when thermal operations are augmented by the possibility of using a catalyst. Catalysts are (non-thermal) ancillary systems which are involved in the overall transformation but are given back unchanged and uncorrelated at the end of the protocol. The ``many second laws'' constraints involve the decrease of a one-parameter family $S_\alpha(\cdot\|\vec{g})$ of relative entropies to the thermal vector $g_n$, with $S_\alpha(\cdot \| \cdot)$ generalising the standard Kullback-Leibler relative entropy: $S_\alpha(\vec{p}\|\vec{g}) \geq S_\alpha(\vec{q}\|\vec{g}) $ for all $\alpha \in \mathbb{R}$. We refer to \cite{brandao2013second} for more details. 

The following considerations and questions arise from our work. First of all, beyond potential subtleties due to infinite dimensions, the many second laws still hold in our scenario, since GTOs are a subset of thermal operations (this was explicitly shown for the single-mode unsqueezed case, since we proved thermo-majorisation and the latter implies the second laws). What is not obvious is, again, that the second laws are sufficient for the existence of a Gaussian catalyst, Gaussian environment and energy-preserving Gaussian unitary mapping the (Gaussian) $\vec{p}$ into $\vec{q}$. Formally, one could define the set of GTOs aided by Gaussian catalysts (``Catalytic GTOs'') and verify the above conjecture. This is a second open question we leave here. Note that allowing for catalysts, while natural from a theoretical perspective, due to the extra control required does not answer the question of the existence of practical setups involving them. 

Finally, one could also look for a Gaussian analogue of the recent result \cite{mueller2017correlating}, especially in the hope that the Gaussian scenario offers `realistic' settings (auxiliary systems and interactions) in which to achieve some of the advantages promised by the correlating thermal machines analysed there, such as the dramatic quashing of unwarranted fluctuations.  

\end{document}